\documentstyle[epsfig,eqsecnum,aps]{revtex}

\newcommand{\mbbm}[1]{{\mbox{\boldmath $#1$}}}

\begin{document}
\draft

\title{Determinant of a new fermionic action on a lattice - (I)}
\author{A. Takami, T. Hashimoto, M. Horibe, and A. Hayashi}
\address{Department of Applied Physics, Fukui University, Fukui 910}
\maketitle

\begin{abstract}
We investigate, analytically and numerically, the fermion determinant of
a new action on a (1+1)-dimensional Euclidean lattice.
In this formulation the discrete chiral symmetry is preserved and the number of
fermion components is a half of that of Kogut-Susskind.
In particular, we show that our fermion determinant is real and positive
for U(1) gauge group under specific conditions,
which correspond to gauge conditions on the infinite lattice.
It is also shown that the determinant is real and positive
for SU(N) gauge group without any condition.
\end{abstract}
\pacs{PACS number(s): 11.15.Ha}

\section{Introduction}

As the Nielsen-Ninomiya theorem \cite{NN} states, we necessarily meet 
the difficulty of so-called fermion doubling
problem when we formulate fermion fields on a lattice.
In practical calculations, Wilson fermions \cite{Wil} have been widely used,
where an additional term which vanishes in the naive continuum limit
is introduced at the expense of the chiral symmetry.
An alternative scheme was proposed by Kogut and Susskind \cite{KS}.
In this scheme the chiral symmetry is maintained as discrete one and
doubler fermions are regarded as fermions in other species.
In $(1+D)$ dimensions, the Kogut-Susskind (KS) formalism describes
a theory with
$2^{\frac{1+D}{2}}$ degenerate quark flavors ($2^{1+D}$ components).

Recently it has been shown that lattice fermionic actions satisfying
the Ginsparg-Wilson relation \cite{GW} may provide a solution of the chirality
problem \cite{Chiral-Problem}.
In these attempts the modified chiral symmetry operator 
is used in stead of  $\gamma_5$. The actions 
are local in the sense 
that the fermionic matrix are bounded  by $Ce^{-\gamma \mid x \mid}$. 
However, it has been proved that actions with the Ginsparg-Wilson 
relation cannot be "ultralocal" \cite{Ultra-Local}.
From the practical point of view, 
the ultralocality (the couplings drop to zero beyond a finite number
of lattice spacings) is also important.
Thus ultralocal fermionic actions with better features than, for example,
the KS action is awaited, though not a final solution of
the chirality problem.  

In the recent papers \cite{Our-Formula1,Our-Formula2},
we proposed a new type of fermionic action on a $(1+D)$-dimensional lattice.
The action is ultralocal and constructed so that fermion fields
satisfy the bosonic
type of dispersion relation. In this sense there are no extra poles
in the propagator. We found that the minimal number of fermion components,
dimensions of the spinor space, is $2^{D-1}$ in the Minkowski case and
$2^D$ in the Euclidean case, 
which should be compared with $2^{1+D}$ of the KS fermion.
Furthermore our action has the discrete chiral symmetry as well. 

It is much of interest to investigate the numerical feasibility of  
our new fermionic action.  When dynamical fermions are included, 
the property of fermion determinants is crucial in numerical 
calculations.
For example, some methods proposed to treat fermionic freedoms rely on 
reality or positivity of the fermion determinants \cite{Some-Methods}.

In this paper we report the analytical and numerical results on the 
fermion determinants of our new action in $(1+1)$ dimensions.
Our main concern is on U(1) gauge group, but some results 
on SU(N) gauge group are also presented.
In the case of U(1) gauge group, we will see that
calculations with specific conditions for temporal link variables
are stable and satisfactory
though results without the conditions are unstable.
The reason why we need those conditions will be discussed in detail.

In Sec.2, we recapitulate our formalism for later convenience.
The analytical and numerical results on fermion determinants will 
be presented in Sec.3, which is followed by the summary.

\section{New fermionic action}

In the previous paper \cite{Our-Formula2},
we proposed a new fermionic action on the Euclidean lattice.
Though the action respects the discrete chiral symmetry
like one in the KS action, fermion fields in this action have $2^D$
components in $(1+D)$ dimensions, which should be compared with $2^{1+D}$ 
in the case of KS fermions. In this section we briefly sketch our formalism
for later convenience.

The action can be written with the fermion matrix $\Lambda$ as
\begin{eqnarray}
	S_f = \sum_{m,n} \psi^\dagger_m \Lambda_{m,n} \psi_n,
	\label{eq:action0}
\end{eqnarray}
where the summation is over lattice points and spinor indices, and our fermion
matrix is defined by
\begin{eqnarray}
	\Lambda = 1 - S_0^\dagger U_E.
	\label{eq:fermi_mat0}
\end{eqnarray}
Here $U_E$ is the Euclidean time evolution operator
and $S_\mu$ is the unit shift operator defined as
\begin{eqnarray}
	S_\mu \psi (x^0, x^1, \ldots, x^\mu, \ldots, x^D)
		 = \psi (x^0, x^1, \ldots, x^\mu+1, \ldots, x^D)
	\qquad	(\mu = 0, 1, \ldots, D).
\end{eqnarray}

We required that the propagator has no extra poles
and found that $U_E$ has the form
\begin{eqnarray}
	U_E = 1 - \sum_{i=1}^D \frac{r_E}{2} \left\{ i X_i
		\left( S_i-S_i^\dagger \right) + \left( 1 - Y_i \right)
		\left( S_i - 2 + S_i^\dagger \right) \right\},
	\label{eq:u_e0}
\end{eqnarray}
where $r_E$ is the ratio of the temporal
lattice constant to the spatial one and $X$'s and $Y$'s,
which are matrices with respect to spinor indices,  
should satisfy the following
algebra:
\begin{eqnarray}
	\left\{
		\begin{array}{ccl}
		\left\{ X_i, X_j \right\}
		& = & \displaystyle{\frac{2}{r_E}} \delta_{ij}, \\
		\left\{ X_i, Y_j \right\}
		& = & 0, \\
		\left\{ Y_i, Y_j \right\}
		& = & 2 \left( \displaystyle{\frac{1}{r_E}} \delta_{ij} + 1 \right),
		\end{array}
	\right.
	\label{eq:algebraXY}
\end{eqnarray}
where $i$ and $j$ run from $1$ to $D$. 
The matrix $2 (\delta_{ij}/r_E+1) $ is
positive definite for any positive $r_E$, therefore $X$'s and
$Y$'s can be assumed hermitian,
\begin{eqnarray}
	X_i^\dagger = X_i, \qquad
	Y_i^\dagger = Y_i.
\end{eqnarray}
The matrices $X$'s and $Y$'s are written in terms of the Clifford algebra
$\Gamma_1, \Gamma_2, \ldots, \Gamma_{2D}$ as
\begin{eqnarray}
	X_i = \displaystyle{\frac{\Gamma_i}{\sqrt{r_E}}}, \qquad
	Y_i = \displaystyle{\sum_{j=1}^D} \alpha_{ij} \Gamma_{D+j},
	\label{eq:XYGamma}
\end{eqnarray}
where
\begin{eqnarray}
	\Gamma_i^\dagger = \Gamma_i, \qquad
	\left\{ \Gamma_i, \Gamma_j \right\} = 2 \delta_{ij}
	\qquad (i,j = 1, \ldots, 2D),
\end{eqnarray}
and $\alpha_{ij}$ are certain real constants.
The dimension of the irreducible representation for $\Gamma$'s is
$2^D$ and accordingly $\psi$ has $2^D$ components.

Now we give some useful properties of the time evolution operator $U_E$.
We can see immediately that
\begin{eqnarray}
	U_E^\dagger = U_E,
\end{eqnarray}
as $X$'s and $Y$'s are hermitian. Since the inverse of $U_E$ is
given by
\begin{eqnarray}
	U_E^{-1} = 1 - \sum_{i=1}^D \frac{r_E}{2}
			\left\{ -i X_i \left( S_i - S_i^\dagger \right)
		+ \left( 1 + Y_i \right)
			\left( S_i - 2 + S_i^\dagger \right) \right\},
	\label{eq:u_inv0}
\end{eqnarray}
we find that $U_E$ is related to its inverse as:  
\begin{eqnarray}
	\Gamma_{2D+1} U_E \Gamma_{2D+1} = U_E^{-1},
	\label{eq:GUG0}
\end{eqnarray}
where
\begin{eqnarray}
	\Gamma_{2D+1} & \equiv & i \Gamma_1, \Gamma_2 \ldots \Gamma_{2D}, \\
	\Gamma_{2D+1}^\dagger = \Gamma_{2D+1}, \qquad \Gamma_{2D+1}^2 = 1,
	& & \qquad \left\{ \Gamma_i, \Gamma_{2D+1} \right\} = 0
	\qquad (i = 1, \cdots, 2D).
\end{eqnarray}

The interaction of the fermion with gauge fields is
introduced by replacing the unit shift operators by covariant ones:
\begin{eqnarray}
	S_\mu \to S_\mu(x) \equiv U_{x, x+\hat{\mu}} S_\mu,
	\label{eq:replace}
\end{eqnarray}
where $\hat{\mu}$ is the unit vector along the $\mu$'th direction,
and $U_{x, y}$ is a link variable connecting sites $x$ and $y$.

The fermion matrix Eq.(\ref{eq:fermi_mat0}) and the
time evolution operator Eq.(\ref{eq:u_e0}) become
\begin{eqnarray}
	\Lambda (x) = 1 - S_0^\dagger(x) U_E(x),
	\label{eq:fermi_mat1}
\end{eqnarray}
and
\begin{eqnarray}
	U_E(x) = 1 - \sum_{i=1}^D \frac{r_E}{2} \left\{ i X_i
		\left(S_i(x) - S_i^\dagger(x) \right)
		+ \left( 1 - Y_i \right)
		\left( S_i(x) - 2 + S_i^\dagger(x) \right) \right\}.
	\label{eq:u_e1}
\end{eqnarray}

With gauge fields coupled to the fermions, however, Eqs.(\ref{eq:u_inv0}) 
and (\ref{eq:GUG0}) do not hold any more in arbitrary dimensions
except for $(1+1)$ dimensions.
In $(1+1)$ dimensions these relations are
\begin{eqnarray}
	U_E^{-1}(x) = 1 - \frac{r_E}{2}
			\left\{ -i X_1 \left( S_1(x) - S_1^\dagger(x) \right)
		+ \left( 1 + Y_1 \right)
			\left( S_1(x) - 2 + S_1^\dagger(x) \right) \right\},
\end{eqnarray}
and
\begin{eqnarray}
	\Gamma_3 U_E(x) \Gamma_3 = U_E^{-1}(x),
	\label{eq:GUG1}
\end{eqnarray}
which are used in the next section.

\section{Analytical and numerical results of our fermion determinant
	in $(1+1)$ dimensions}

\subsection{Reality of the determinant}

In this section we show the analytical and  numerical results of
our fermion determinant in the $(1+1)$-dimensional case.

First, we investigate the determinant of the unit shift operator $S_0(x)$
in $(1+D)$ dimensions.
Let us denote the link variables by $U_{x, x+\hat{\mu}}$ which live on the
links connecting two neighboring lattice sites $x$ and $x+\hat{\mu}$. 
Now we consider the case of the Abelian gauge group U(1),
and they can be written in the form
\begin{eqnarray}
    U_{x, x+\hat{\mu}} = e^{i \theta_\mu(x)},
\end{eqnarray}
where $\theta_\mu(x)$ is restricted to the compact domain [0, 2$\pi$).

Then, $S_0(x)$ can be constructed as a
$N_T N_{\mbbm{x}} N_s \times N_T N_{\mbbm{x}} N_s$ matrix 
where $N_T$ and $N_{\mbbm{x}}$ is the size of the lattice in the Euclidean time
and the spatial directions, respectively, and 
$N_s$ is the number of spinor components.
$S_0(x)$ is diagonal in the spatial and spinor space and consists of
block $N_T \times N_T$ matrices $Q(\mbbm{x})$ belonging to 
the Euclidean time space. 
Thus we can write the determinant of $S_0(x)$ as follows:
\begin{eqnarray}
	\det S_0(x) = \left|
	\begin{array}{cccccc}
		Q(1)    &        &      &        &               &        \\
				& \ddots &      &        &               &        \\
				&        & Q(2) &        &               &        \\
				&        &      & \ddots &               &        \\
				&        &      &        & Q(N_\mbbm{x}) &        \\
    		    &        &      &        &               & \ddots \\
	\end{array}
	\right|,
\end{eqnarray}
where $Q(\mbbm{x})$ for each $\mbbm{x}$ appear $N_s$ times
as $S_0(x)$ is unity in the spinor space.

The determinant of $Q(\mbbm{x})$ is easily calculated by noticing its matrix
form:
\begin{eqnarray}
	\det Q(\mbbm{x}) & = & \left|
		\begin{array}{ccccc}
			0 & U_{(1, \mbbm{x}),(2, \mbbm{x})} & & &         \\
			& 0 & U_{(2, \mbbm{x}),(3, \mbbm{x})} & &         \\
			& & \ddots &                     \ddots &         \\
			& & & 0 & U_{(N_{T-1}, \mbbm{x}),(N_T, \mbbm{x})} \\
			-U_{(N_T, \mbbm{x}),(1, \mbbm{x})} & & & & 0      \\
		\end{array}
	\right|
	\label{eq:detQ}
	\\
		& = & (-)^{N_T-1} U_{(1, \mbbm{x}),(2, \mbbm{x})}
				U_{(2, \mbbm{x}),(3, \mbbm{x})} \ldots
				\left( -U_{(N_T, \mbbm{x}),(1, \mbbm{x})} \right) \nonumber \\
		& = & (-)^{N_T}
			\exp \left\{ i \sum_{t=1}^{N_T} \theta_0(t, \mbbm{x}) \right\}.
\end{eqnarray}
We used the anti-periodic boundary condition for the time direction, which 
is represented by a negative sign in the lower-left corner of the matrix $Q$. 

As a result, in $(1+D)$ dimensions we find
\begin{eqnarray}
	\det S_0(x) & = & \prod_{\mbbm{x}}
			\left[ (-)^{N_T} \exp \left\{ i \sum_{t=1}^{N_T}
			\theta_0(t,\mbbm{x}) \right\} \right]^{N_s} \nonumber \\
		& = & \exp \left\{ i N_s \sum_t \sum_{\mbbm{x}}
			\theta_0(t,\mbbm{x}) \right\} = e^{i\alpha},
	\label{eq:detS0}
\end{eqnarray}
where the angular variable $\alpha$ is defined by
\begin{eqnarray}
	\alpha = N_s \sum_t \sum_{\mbbm{x}} \theta_0(t,\mbbm{x}).
	\label{eq:alpha_res}
\end{eqnarray}

From now we assume $D = 1$ and show that the determinant
of the Euclidean time evolution operator $U_E(x)$ is unity.
By Eq.(\ref{eq:GUG1}) we easily find $\det U_E(x) = \pm 1$.
Since $X$'s and $Y$'s obey Eq.(\ref{eq:XYGamma}) and
in the limit $r_E \to 0$
\begin{eqnarray}
    r_E X_i \longrightarrow 0, \qquad
    r_E Y_i \longrightarrow 0,
\end{eqnarray}
$\det U_E(x)$ is equal to $1$ at $r_E = 0$.
From the continuity of $\det U_E(x)$ with respect to $r_E \ge 0$, we conclude
\begin{eqnarray}
    \det U_E(x) = 1,
    \label{eq:det_u_e_1}
\end{eqnarray}
in the $(1+1)$-dimensional case.

Collecting the above results, the relations (\ref{eq:GUG1}) and
(\ref{eq:det_u_e_1}), and the hermiticity of $U_E(x)$,
we can obtain a relation on the phase of our fermion determinant
in $(1+1)$ dimensions as follows:
\begin{eqnarray}
	\left( \det \Lambda (x) \right)^\ast
		& = & \det \left( 1 - S_0^\dagger(x) U_E(x) \right)^\dagger
			= \det \left( 1 - U_E(x) S_0(x) \right) \nonumber \\
		& = & \det U_E(x) \det S_0(x)
			\det \left( S_0^\dagger(x) U_E(x)^{-1} - 1 \right) \nonumber \\
		& = & (-)^{\mbox{dim. of $\Lambda$}} \det S_0(x) \det \Lambda(x).
\end{eqnarray}
Since the number of components of our fermion is $N_s = 2$
and consequently the fermion matrix $\Lambda(x)$
always has an even number of dimensions, we get
\begin{eqnarray}
	\left( \det \Lambda(x) \right)^\ast = e^{i \alpha} \det \Lambda(x).
\end{eqnarray}

Accordingly in the case of $(1+1)$ dimensions with U(1) gauge fields,
our fermion determinant is real under the condition
\begin{eqnarray}
	\alpha = 2 \pi n \qquad \mbox{($n$: integer)},
	\label{eq:a-res}
\end{eqnarray}
where $\alpha$ is defined in Eq.(\ref{eq:alpha_res}).
The condition specified by Eq.(\ref{eq:a-res}) will be called "GT-condition"
in this paper, where the temporal link variables are globally constrained as
$\sum_x \theta_0(x) = \pi n$ and on the infinite lattice the condition
is achieved by a gauge transformation.
We also define "T-condition" as $\theta_0(x) = 0$, which corresponds to
the temporal gauge condition on the infinite lattice.

At a glance, it seems strange that the determinant has $\alpha$-dependence,
because $\alpha$ can be gauged away.
However, this is not true for a finite lattice,
since all gauge transformations on the infinite lattice are not allowed
on a finite lattice with periodic boundary conditions.
Thus, the dependence comes from finiteness of the lattice.
For example, $\sum_t \theta_0(t, \mbbm{x})$ is invariant
under any gauge transformation on a finite lattice with periodic
boundary conditions. The temporal gauge condition is certainly a 
consistent gauge in an infinite system, but not in a finite system.
However, this restriction is just an artifact
arising when we try to approximate an infinite system by a finite one.
We could think in the following way. Given an infinite system, we impose
the temporal gauge condition, which is legitimate. Then we take an finite
sub-system to approximate the original total system.  
The difference is due to size and boundary effects and would disappear in 
the infinite volume limit. 
The same is also true for the GT-condition. 
It should be noted that the fermion determinant is invariant under 
gauge transformations on a finite lattice,
but the determinant does depend on our "gauge conditions":
the T- or the GT-condition.

\begin{figure}[htbp]
	\begin{center}
		\epsfig{file=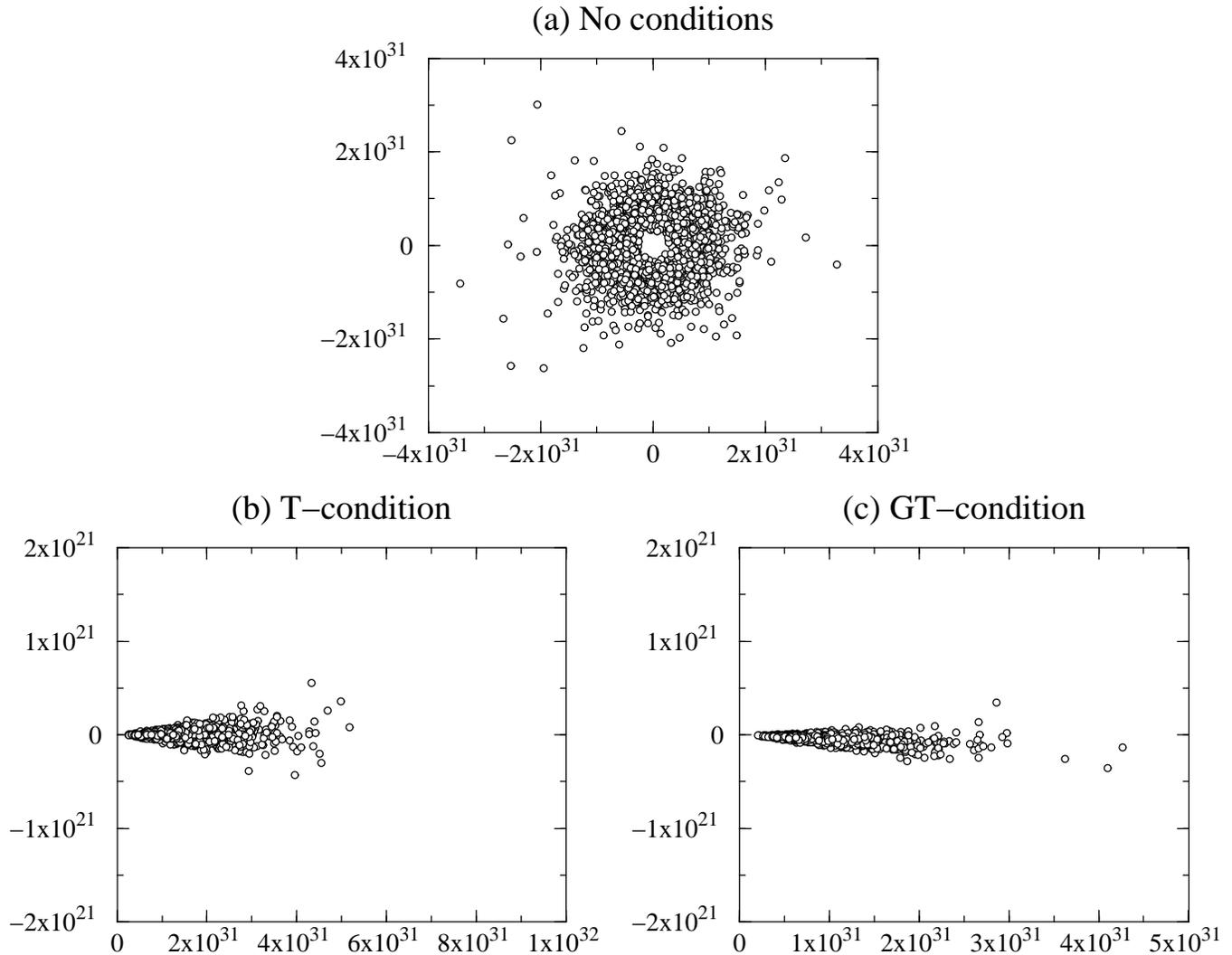}
	\end{center}
	\caption{The distributions in the complex plane of our fermion determinants
		for each configuration of 2000 Monte Carlo iterations
		after getting good enough equilibrium, i.e. after 2000 iterations,
		on a $8 \times 8$ lattice at $\beta = 1.5$.
		Plots (a),(b) and (c) correspond to under no conditions,
		the T- and the GT-condition, respectively.}
	\label{fig01}
\end{figure}

\newpage

Now, we will give the numerical results
in the $(1+1)$-dimensional U(1) gauge theory with our fermion action.
In the following numerical simulations link variables are updated by
the Metropolis method and determinants are calculated by the LU decomposition.
So there are no systematic errors in the determinants.
The way to generate a sequence of configurations in the Monte Carlo computation
is as follows.
In the T-condition, we fix $U_0(x) = 1$ everywhere.
After this, all temporal link variables are not changed
and we only update the link variables of spatial direction
using the Metropolis method.
In the GT-condition,
we prepare a configuration satisfying $\sum_x \theta_0(x) = \pi n$.
Then a temporal link variable $e^{i \theta_0(x)}$
is replaced by $e^{i \theta_0(x) + i \chi}$, where $\chi$ is a random number
between $- \pi$ and $\pi$.
At the same time another temporal link variable, which is chosen randomly,
is multiplied by $e^{- i \chi}$.
And we use the above configuration as the trial one
in a Metropolis acceptance test.
It can be shown that this procedure satisfies
the micro-reversibility requirement,
and therefore also the detailed balance condition.

In Fig.\ref{fig01}(a) - \ref{fig01}(c) we show the distribution of the fermion
determinants in the complex plane for some $\beta$. The statistics is
$2000$ thermalizations and $2000$ measurements.
In Fig.\ref{fig01}(a), no conditions are imposed on temporal link variables.
Fig.\ref{fig01}(b) is the result in the T-condition.
In Fig.\ref{fig01}(c), the GT-condition with $n = 0$ is imposed,
namely the condition $\alpha = 0$ in Eq.(\ref{eq:alpha_res})
is always kept in Monte Carlo updates.
The distribution in Fig.\ref{fig01}(a) has a doughnut-like structure
with a center around the origin, so that
we can expect that the convergence for any observation is very poor.
On the other hand, Fig.\ref{fig01}(b) and \ref{fig01}(c) show that
the determinants in the T- and the GT-condition are real
as expected and also positive.
The distribution in the imaginary direction is due to numerical errors
(note the difference in the scales of real and imaginary parts).
The positivity of determinants is important from the numerical point of view
and will be discussed later.
 
To test the convergence in the above three types of conditions,
we measured the expectation value of a plaquette value
as a function of $\beta$ using the following formula:
\begin{eqnarray}
	\langle W_p \rangle
		= \frac{\displaystyle{\int dU d\psi d\psi^\dagger W_p e^{-S_g - S_f}}}
			{\displaystyle{\int dU d\psi d\psi^\dagger e^{-S_g - S_f}}}
		= \frac{\displaystyle{\int dU W_p \det \Lambda(x) e^{-S_g}}}
			{\displaystyle{\int dU \det \Lambda(x) e^{-S_g}}}
		= \frac{\displaystyle{\langle W_p \det \Lambda(x) \rangle}_0}
			{\displaystyle{\langle \det \Lambda(x) \rangle}_0}.
	\label{eq:pla}
\end{eqnarray}
Here $S_g$ is the usual action for link fields,
and $d\psi$, $d\psi^\dagger$ and $dU$ stand for
$\prod_{x, \alpha} d\psi_{\alpha}(x)$,
$\prod_{x, \alpha} d\psi^\dagger_{\alpha}(x)$ and
$\prod_{x, \mu} dU_{x, x+\hat{\mu}}$, respectively.

\begin{figure}[htbp]
    \begin{center}
        \epsfig{file=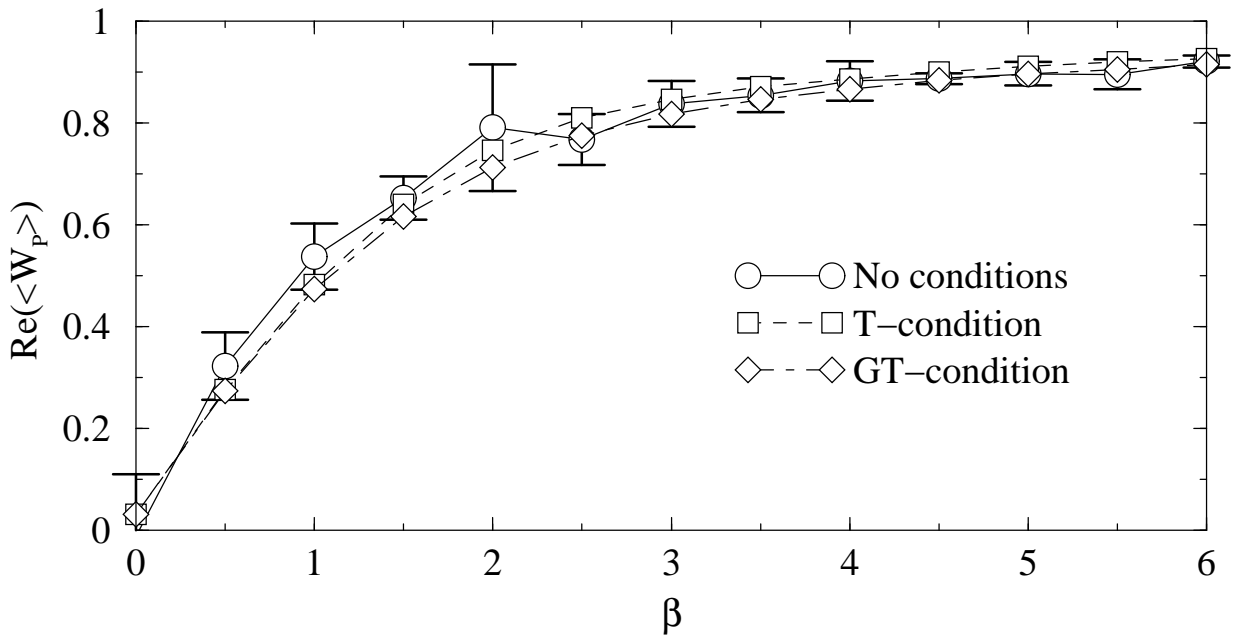}
    \end{center}
    \caption{The real part of the averaged plaquette vs $\beta$
        on a $8 \times 8$ lattice.}
    \label{fig02}
\end{figure}

\newpage

In Fig.\ref{fig02} we can see that the expectation values
in the T- and the GT-condition display gentle curves,
while one without any conditions is spiky especially for small $\beta$.
Here, the plotted points are the average over $10$ results,
each of which is obtained by $2000$ Monte Carlo iterations at each $\beta$.
And the error bars are evaluated by using the standard deviation
of the $10$ results.
If the error bars are not displayed, they are not visible at this scale.

\begin{figure}[htbp]
    \begin{center}
        \epsfig{file=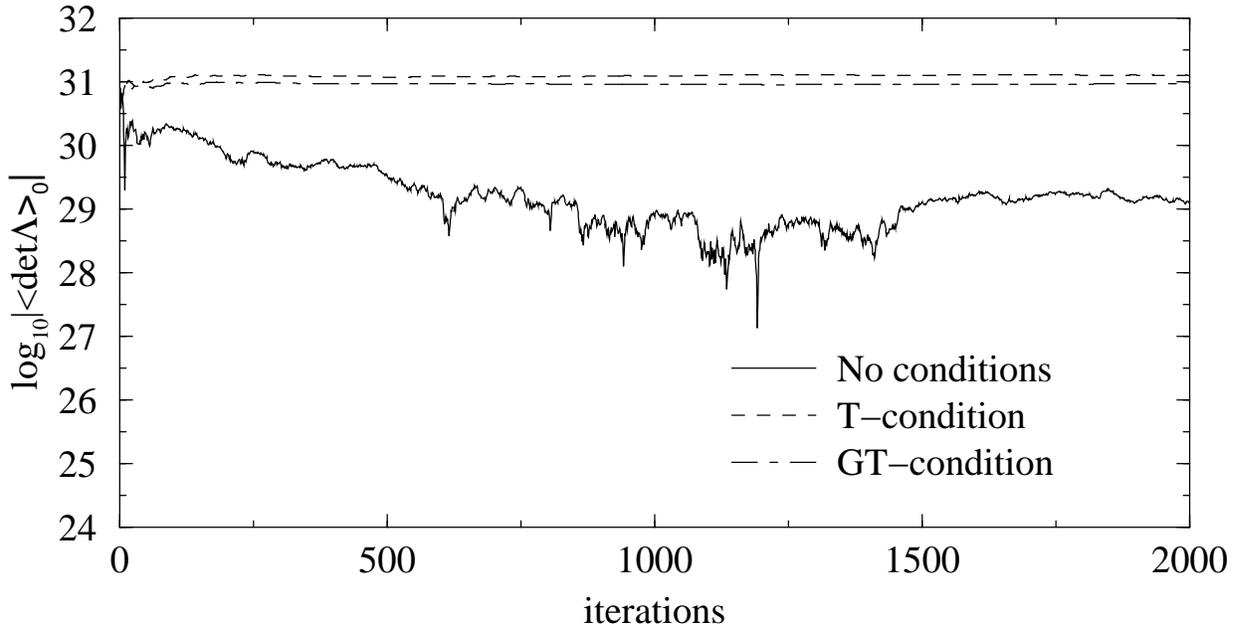}
    \end{center}
    \caption{The absolute value of the averaged fermion determinant
        vs Monte Carlo iterations. The data were computed
        using the same configurations as in Fig.1(a) - 1(c)}
    \label{fig03}
\end{figure}

The poorness of the convergence without any conditions can be traced back 
to the behavior of the denominator in Eq.(\ref{eq:pla}).
The expectation value
$\langle \det \Lambda(x) \rangle_0$ not only takes complex values
but also suddenly increases or decreases during sampling.
Fig.\ref{fig03} shows the absolute value  of the averaged fermion determinant
as a function of the update iterations for each condition.
We find that the convergence in the case of no conditions is ill
while in the cases of other two conditions very fine.
Generally the more iterations are expected to improve the convergence. 
Without any conditions, however, this is not the case since 
the expectation value of the determinant must be in the empty hole of
a doughnut-like structure in Fig.\ref{fig01}(a).
Thus it is difficult to improve the convergence for
any expectation value within a reasonable number of iterations.

\begin{figure}[htbp]
	\begin{center}
		\epsfig{file=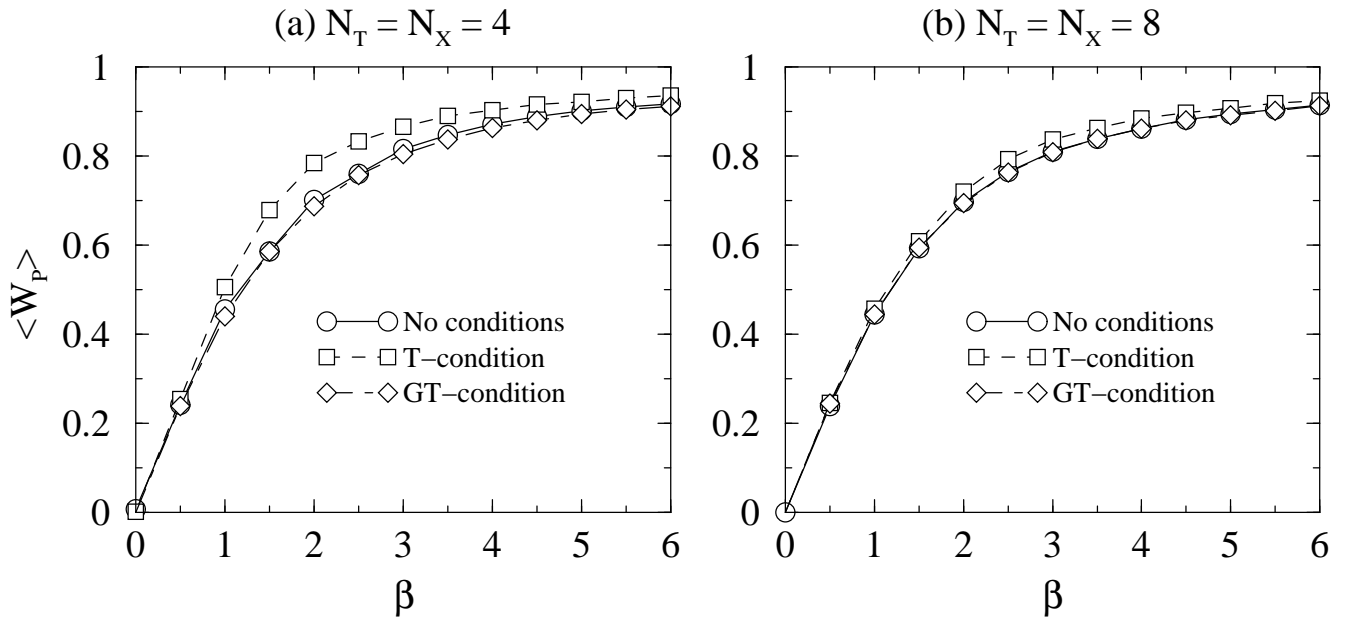}
	\end{center}
	\caption{The averaged plaquette vs $\beta$ on a (a) $4 \times 4$
		and (b) $8 \times 8$ lattice.}
	\label{fig04}
\end{figure}

Now what is the difference between the T- and the GT-condition?
We expect that 
the GT-condition is superior to the T-condition
since the former condition is much weaker condition than the latter. 
To see this, we show in Figs.\ref{fig04}(a) and \ref{fig04}(b)
the averaged plaquette value
in the pure U(1) gauge theory as a function of $\beta$
for three types of conditions.
We find the line without any conditions and one with the GT-condition
are very close to each other, but the line calculated in the T-condition
is slightly upper than those.
As expected, this difference comes from the size effects
in each condition and obviously tends to decrease
as the lattice size becomes larger.
Thus we conclude that the GT-condition has better feature on a finite lattice:
the fermion determinant is positive and the finite-size effects are smaller.

\subsection{Spectrum of the fermion matrix}

Before discussing the positivity of the fermion determinant,
we study the spectrum of $\Lambda(x)$.
First, we introduce a discrete rotational symmetry in the complex plane 
for eigenvalues of $S_0^\dagger(x) U_E(x)$.
Defining $(T)_{t, t^\prime} = \exp (i 2 \pi t / N_T) \delta_{t, t^\prime}$,
we find
\begin{eqnarray}
	T S_0^{\dagger}(x) U_E(x) T^{-1} = \exp \left(
		\displaystyle{i \frac{2 \pi}{N_T}} \right)
		S_0^{\dagger}(x) U_E(x).
\end{eqnarray}
This relation implies that if $\lambda$ is some eigenvalue of
$S_0^\dagger(x) U_E(x)$, then $\exp \left(i 2 \pi / N_T \right) \lambda$ is
also its eigenvalue.
Next using the following relation
\begin{eqnarray}
    \det \left( U_E(x) S_0(x) - \lambda^\ast \mbox{\bf{1}} \right)
        & = & \det \left( U_E(x) S_0(x) \right)
            \det \left( \mbox{\bf{1}} - \lambda^\ast S_0^{\dagger}(x)
            U_E(x)^{-1} \right) \nonumber \\
        & = & \det \left( U_E(x) S_0(x) \right)
			\det \left( \mbox{\bf{1}} - \lambda^\ast S_0^{\dagger}(x)
			U_E(x) \right),
\end{eqnarray}
we rewrite the eigenvalue equation
\begin{eqnarray}
	\det \left( S_0^{\dagger}(x) U_E(x) - \lambda \mbox{\bf{1}} \right) = 0
	\label{eq:det_S0dUE}
\end{eqnarray}
into the form:
\begin{eqnarray}
	\det \left( S_0^{\dagger}(x) U_E(x)
		- {\lambda^\ast}^{-1} \mbox{\bf{1}} \right) = 0.
	\label{eq:det_S0dUE_inv}
\end{eqnarray}
Accordingly, from Eqs.(\ref{eq:det_S0dUE}) and (\ref{eq:det_S0dUE_inv})
if $\lambda$ is some eigenvalue of $S_0^\dagger(x) U_E(x)$,
then ${\lambda^\ast}^{-1}$ is also its eigenvalue.

\begin{figure}[htbp]
	\begin{center}
		\epsfig{file=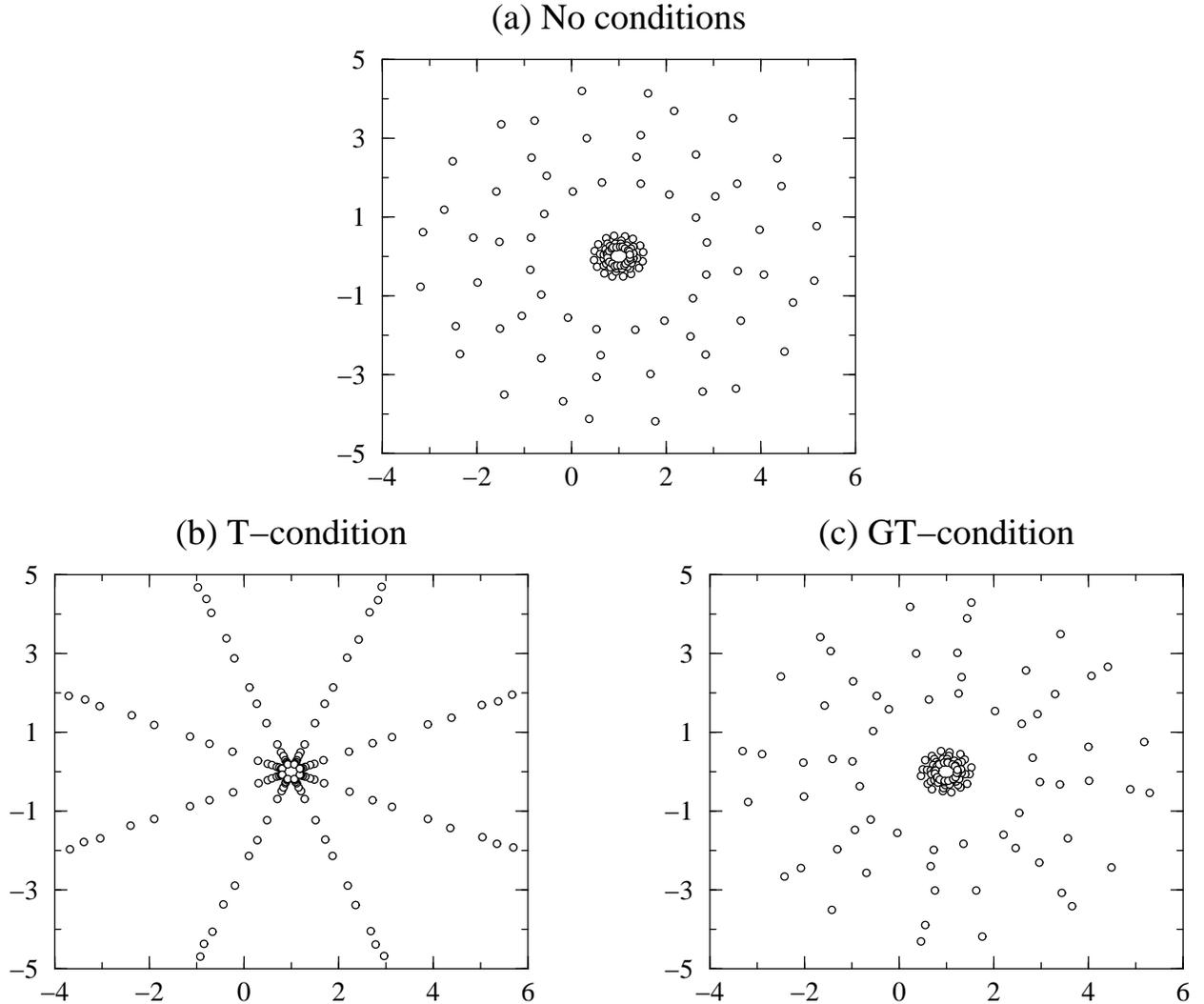}
	\end{center}
	\caption{The spectrum of $\Lambda(x)$ in the complex plane
		for a $8 \times 8$ lattice in the case of (a) no conditions,
		(b) the T- and (c) the GT-condition, respectively.}
	\label{fig05}
\end{figure}

Let us look at the numerical results of the spectrum of
our fermion matrix $\Lambda(x) = 1 - S_0^{\dagger}(x) U_E(x)$
in the $(1+1)$-dimensional U(1) theory.
Figs.\ref{fig05}(a) - \ref{fig05}(c) display the spectrum
in the case of no conditions, the T- and GT-condition, respectively.
From the figures we confirm the two properties of the spectrum discussed
above for each condition.
The spectrum in the GT-condition is so similar to the one
with no conditions that one couldn't distinguish them at a glance.
It is interesting that the determinant is real only in the former case.

Now, using above two properties for $S_0^\dagger(x)U_E(x)$, we discuss
the positivity of our fermion determinant. We will give a plausible
reason for the positivity not a complete proof.

First we write the determinant as follows:
\begin{eqnarray}
	& & \det \left( 1 - S_0^\dagger(x) U_E(x) \right) \nonumber \\
	& & \;\;\; = \left( 1 - \lambda_1 \right)
		\left( 1 - \frac{1}{{\lambda_1}^\ast} \right)
		\left( 1 - \lambda_2 \right)
		\left( 1 - \frac{1}{{\lambda_2}^\ast} \right) \cdots \nonumber \\
	& & \;\;\; = \frac{1}{ \left( - \right)^{N_T N_X} }
		\frac{ \left( 1 - \lambda_1 \right)
		\left( 1 - {\lambda_1}^\ast \right) }{{\lambda_1}^\ast}
		\frac{ \left( 1 - \lambda_2 \right)
		\left( 1 - {\lambda_2}^\ast \right) }{{\lambda_2}^\ast} \cdots
	\label{eq:det_from_eigen}
\end{eqnarray}
In Eq.(\ref{eq:det_from_eigen}), the denominator must be real,
since the numerator is positive
and $\det\Lambda(x)$ is real under the T- or the GT-condition as shown before.
The denominator is a continuous function of
the background configuration and furthermore it is never vanished,
because $S_0^\dagger(x) U_E(x)$
has always an inverse. Once the denominator is positive for some background
configuration, it keeps on having a positive value for any configuration, if
one configuration can be transformed continuously to the other
within the T- or the GT-condition.

\newpage

Next, we show that our fermion determinant is positive when all link
variables are set unity.
In this case $S_0^\dagger U_E$ can be easily diagonalized
in the momentum space.
The eigenvalues are expressed by the eigenvalues ${\lambda_U}_m$ of $U_E$ as
\begin{eqnarray}
	{\lambda_U}_m \exp \left( i \frac{2n+1}{N_T} \pi \right)
		\qquad (n = 0, \ldots, N_T-1).
\end{eqnarray}
Therefore
\begin{eqnarray}
	\det \Lambda = \prod_{m,n}
		\left( 1 - {\lambda_U}_m
			\exp \left( i \frac{2n+1}{N_T} \pi \right) \right)
\end{eqnarray}
is clearly positive, as ${\lambda_U}_m$ can be shown positive.

The configuration with all link variables unity clearly satisfies
the T- and the GT-condition.
This seems to complete our proof for the positivity of the determinant.
It should be noticed that we might be faced with configurations where
some $\lambda$ with $\left| \lambda \right| = 1$ is not degenerated in spite of
the symmetry $\lambda$ and ${\lambda^\ast}^{-1}$.
In this case the relation Eq.(\ref{eq:det_from_eigen}) does not hold
and our proof fails.
However, such configurations are very hard to happen.

\subsection{SU(N) gauge group}

\begin{figure}[htbp]
    \begin{center}
        \epsfig{file=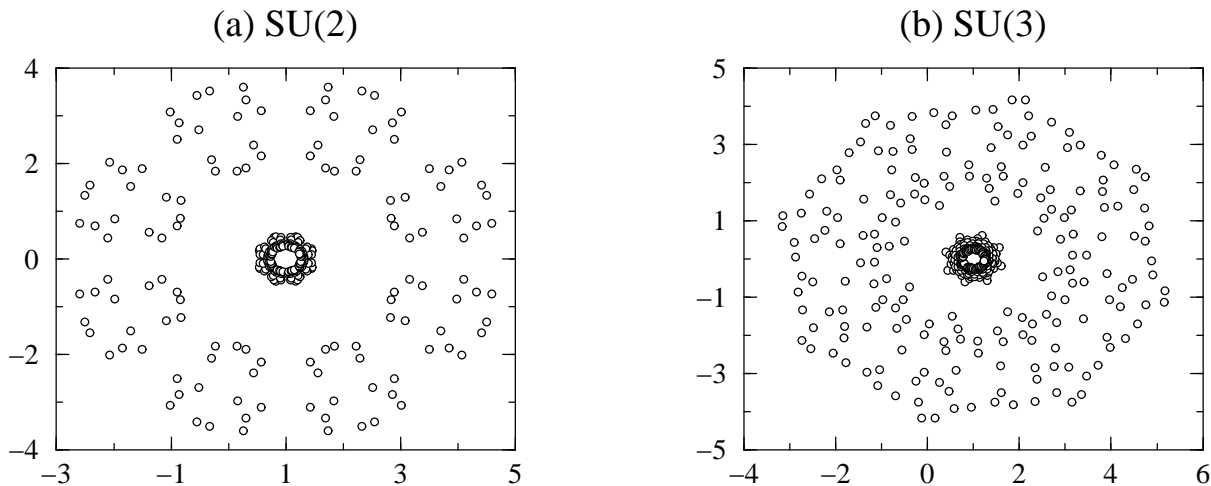}
    \end{center}
    \caption{The spectrum of $\Lambda(x)$ in the complex plane
		for a $8 \times 8$ lattice. Plots (a) and (b) correspond
        to SU(2) and SU(3) gauge groups, respectively.}
	\label{fig06}
\end{figure}

\begin{figure}[htbp]
    \begin{center}
        \epsfig{file=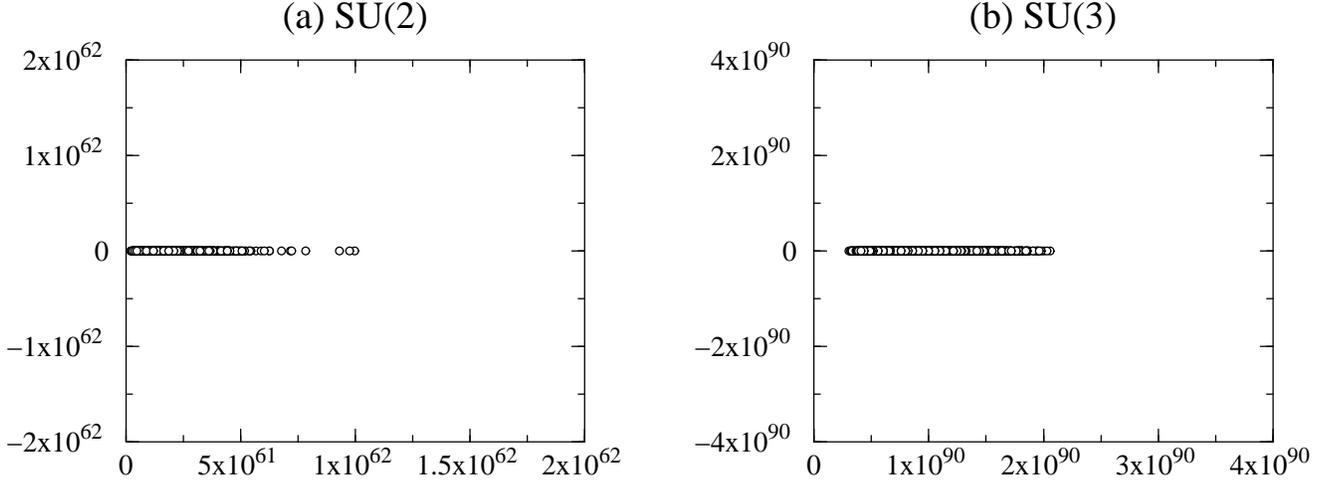}
    \end{center}
    \caption{The distributions of our fermion determinant in the complex plane
		for each configuration of 2000 Monte Carlo iterations
		on a $8 \times 8$ lattice
        at $\beta = 2.0$. Plots (a) and (b) correspond to SU(2) and
        SU(3) gauge groups, respectively.}
	\label{fig07}
\end{figure}

When the gauge group is SU(N), we can also make similar discussion
to the U(1) gauge group case
and find the positivity of our fermion determinant.
In this case, from Eq.(\ref{eq:detQ}) we immediately see
\begin{eqnarray}
    \det Q(\mbbm{x}) & = & \pm \left| U_{(1,\mbbm{x}),(2,\mbbm{x})} \right|
        \left| U_{(2,\mbbm{x}),(3,\mbbm{x})} \right| \ldots
            \left| U_{(N_T-1,\mbbm{x}),(0,\mbbm{x})} \right| \nonumber \\
        & = & \pm 1,
\end{eqnarray}
so we find $\det S_0(x) = 1$ without any conditions.
Consequently, the determinant of our fermion matrix is
always real in the case of SU(N) gauge fields.
And we find that our fermion determinant is positive
in the $(1+1)$-dimensional SU(N), since the two symmetries of the spectrum
of $\Lambda(x)$ discussed above are also satisfied.
In fact we can confirm the symmetries
from the spectra shown in Figs.\ref{fig06}(a) and \ref{fig06}(b) 
for SU(2) and SU(3), respectively.
And, in Figs.\ref{fig07}(a) and \ref{fig07}(b), we display
the numerical results
for the distribution of our fermion determinant in the $(1+1)$-dimensional
SU(2) and SU(3) gauge theories.
The results show that fermion determinants are real and positive
in both cases.

We conclude that our fermion determinant $\det \Lambda(x)$
in (1+1)-dimensions is real and positive for U(1) gauge group
under the T- or the GT-condition and for SU(N) gauge group
without any conditions,
though we have only a plausible reason for the positivity.

\section{Discussion and summary}

When we applied our new action without any conditions to U(1) gauge theory
on a lattice,
we were faced with the problem of convergence in Monte Carlo simulation.
In this note, we showed that we could avoid this problem by
imposing the T- or the GT-condition.
In order to make this situation clear, as an example,
we study the propagator of the fermi field,
\begin{eqnarray}
	\langle \psi(y) \psi^\dagger(y^\prime) \rangle
		= \frac{\displaystyle{\int d\psi d\psi^\dagger dU \psi(y)
			\psi^\dagger(y^\prime) e^{-S_g-S_f}}}
		{\displaystyle{\int d\psi d\psi^\dagger dU e^{-S_g-S_f}}}.
	\label{propa}
\end{eqnarray}
Integrating in Eq.(\ref{propa}) 
with respect to $\psi(x)$ and $\psi^\dagger(x)$ we have
\begin{eqnarray}
	\langle \psi(y) \psi^\dagger(y^\prime) \rangle
		= \frac{\displaystyle{\int dU
			\det \left( 1 - S_0^\dagger(x) U_E(x) \right)
			\left( 1 - S_0^\dagger(x)
				U_E(x) \right)_{y, y^\prime}^{-1} e^{-S_g}}}
		{\displaystyle{\int dU
			\det \left( 1 - S_0^\dagger(x) U_E(x) \right) e^{-S_g}}}.
	\label{propa2}
\end{eqnarray}
We introduce new variables $\Theta$ and $\tilde{\theta}_0(x)$
instead of link variables $U_{x, x+\hat{0}} = e^{i \theta_0(x)}$,
\begin{eqnarray}
	\theta_0(x) = \Theta + \tilde{\theta}_0(x),
\end{eqnarray}
and using the following equation
\begin{eqnarray*}
	& & \int \prod_x d\tilde{\theta}_0(x) d\Theta
        \delta \left( \sum_x \tilde{\theta}_0(x) \right) \\
    & & \;\;\; = \int \prod_x d\theta_0(x) d\Theta
        \delta \left( \sum_x \theta_0(x) - N_T N_X \Theta \right) \\
	& & \;\;\; = \frac{1}{N_T N_X} \int \prod_x d\theta_0(x),
\end{eqnarray*}
we get
\begin{eqnarray}
	& & \langle \psi(y) \psi^\dagger(y^\prime) \rangle \nonumber \\
    & & \;\;\; = \frac{\displaystyle{\int d\tilde{U} d\Theta
		\delta \left( \sum_x \tilde{\theta}_0(x) \right)
		\det \left( 1 - e^{i\Theta} {\tilde{S}_0(x)}^\dagger U_E(x) \right)
		\left( 1 - e^{i\Theta}
			{\tilde{S}_0(x)}^\dagger U_E(x) \right)_{y, y^\prime}^{-1}
		e^{-\tilde{S}_g}}}
		{\displaystyle{\int d\tilde{U} d\Theta
		\delta \left( \sum_x \tilde{\theta}_0(x) \right)
		\det \left( 1 - e^{i\Theta} {\tilde{S}_0(x)}^\dagger U_E(x) \right)
		e^{-\tilde{S}_g}}},
	\label{propa3}
\end{eqnarray}
where the tilde represents replacing $\theta_0(x)$ by
$\tilde{\theta}_0(x)$.

The phase of determinant
$\det (1 - e^{i\Theta} {\tilde{S}_0(x)}^\dagger U_E(x))$ takes any value
between $0$ and $2\pi$ as is shown in Fig.\ref{fig01}(a).
The summation of $\det (1 - e^{i\Theta_n} {\tilde{S}_0(x)}^\dagger U_E(x))$
over a sequence $\Theta_n (n = 1, 2, \ldots)$ which is chosen
at random are canceled out accidentally, then the denominator
of Eq.(\ref{propa3}) becomes very small.
This is the origin of unstable behavior in Monte Carlo simulation.
Indeed in Sec.3, under the T- or the GT-condition
which ensures that the variable $\Theta$ is fixed zero,
we proved the determinant $\det (1 - S_0^\dagger(x) U_E(x))$
in the $(1+1)$-dimensional lattice to be real
for all configurations of link variables and to be positive for most ones.

We have another reason for fixing $\Theta$, namely, imposing some condition
like the T- or the GT-condition.
Since the integrand of the numerator in Eq.(\ref{propa3}) is
equal to the cofactor of matrix
$(1 - e^{i\Theta} {\tilde{S}_0(x)}^\dagger U_E(x))$ whose
elements are linear functions of $e^{i\Theta}$,
it is a polynomial of $e^{i\Theta}$.
After we integrate it with respect to $\Theta$,
all terms of this polynomial vanish besides constant terms.
For the denominator of Eq.(\ref{propa3}) we can say the same thing
as the above, thus we have
\begin{eqnarray}
	\langle \psi(y) \psi^\dagger(y^\prime) \rangle
		= \mbox{\bf{1}}_{y, y^\prime},
\end{eqnarray}
which is an undesired result.
Contrarily, if we choose some condition, like the T- or the GT-condition,
we may avoid this trouble.

When we try applying this fermion to the SU(N) lattice gauge theory,
we can expect that there is no necessity for imposing some condition
because the element like $e^{i\Theta}$ does not belong to the group SU(N).
In fact, it is shown that the fermion determinant is real
for all configurations and that it is positive for most configurations.
And the numerical simulation for SU(2) and SU(3) gauge theory shows that
the fermion determinants are real and positive.
We will discuss the application of our fermions to SU(N) lattice gauge theory
in higher dimensions elsewhere.

\section*{Acknowledgments}
We would like to thank Prof. H. Yamamoto for very fruitful discussions
in the early stages of this work.


\end{document}